\documentclass[aps,twocolumn,floatfix,showpacs,amssymb]{revtex4}
\usepackage{graphicx,bm}
\usepackage{amsmath}
\begin{document}
\title{Implicit ladder summation in the Hartree-Fock-Bogoliubov approach}
\author{Ludovic Pricoupenko}
\affiliation
{
Laboratoire de Physique Th\'{e}orique de la Mati\`{e}re Condens\'{e}e, 
Universit\'{e} Pierre et Marie Curie and CNRS, 4 place Jussieu, 75252 Paris, France.
}
\date{\today}
\begin{abstract}
The fully variational Hartree Fock Bogoliubov approach for bosons is studied in the limit of zero range forces in two- and three-dimensions. The equation of state obtained in two-dimensions is expressed in a parametric form. It is shown that the $\Lambda$ potential permits to perform an implicit summation of the  ladder diagrams without leaving the variational scheme, restoring thus the consistency of this approximation.
\end{abstract}
%
\maketitle

\section{Introduction}

Modeling interatomic forces by a point-like isotropic potential appeared fruitful in the context of ultracold atoms as a way to study  the dilute and low temperature regime for bosons or two-spin component fermions. In this respect, the Bethe Peierls model where pairwise interactions between particles are replaced by contact conditions is especially relevant to finding non trivial properties in the two- and few-body problems, like confinement induced resonances \cite{Ols98,Pet00,Pet01} or also in universal collisional properties as in the four-fermion problem \cite{Pet04a}. Nevertheless the Bethe Peierls formalism is not adapted to usual formalisms of the many-body problem where the interaction is described by a standard two-body vertex \cite{Abr89,Bla86,Fet03}. The Fermi-Huang pseudo-potential which encapsulates the Bethe Peierls contact condition is a way to implement directly the zero-range limit in a perturbative approach of the many-body problem and was  successfully used in the description of the dilute Bose gas \cite{Hua57}. However when applied to the fully variational Hartree Fock Bogoliubov (HFB) approach, the Fermi-Huang pseudo-potential describes a molecular Bose gas instead of the usual atomic phase \cite{Pri00}. This problem is solved by using the $\Lambda$ potential which in exact treatments, provides a description strictly equivalent to the Bethe Peierls model, while it permits to improve approximate formalisms \cite{Ols01}. This potential is a function of the free parameter $\Lambda$ which offers the opportunity to suppress the gap in the HFB description of the homogeneous Bose gas, making thus the formalism compatible with the Hugenholtz Pines theorem \cite{Hug59}. In Ref.~\cite{Pri04b}, the formalism was applied to the two-dimensional (2D) Bose gas. It provides an equation of state (EOS) which coincides with the result of Popov in the dilute limit \cite{Pop72,Mor03} while giving at finite areal density an approximation  which compares nicely with exact Monte-Carlo computations and perturbative results \cite{Pil05,Ast09,Mor09}. Moreover the formalism solves the problem of the unexpected stabilization of a vortex in a non rotating trap, a phenomenon which occurs when HFB equations are truncated by neglecting the effect of the anomalous average of the noncondensate field operator \cite{Gri96,Iso99}. However, the derivation of the HFB equations using the $\Lambda$ potential is not straightforward because the ansatz made for the density matrix doesn't belong to the correct Hilbert space (it doesn't support all the nontrivial singularities linked to the Bethe Peierls contact condition).

The present paper gives a simple derivation of the fully variational HFB equations for the three-dimensional (3D) and the 2D Bose gas by using  the ${\Lambda-\epsilon}$ potential introduced in Ref.~\cite{Pri11}. This potential enables to relax the strict zero range limit for the variational calculation while keeping the structure of the original $\Lambda$ potential and the exact zero range limit is achieved afterward. This way, HFB equations are obtained without any approximation following closely the standard and general variational scheme described for example in Ref.~\cite{Bla86}. Emphasis is put on the 2D case where a parametric form of the non trivial EOS is derived. Finally, the ladder approximation of the Bethe Salpeter equation is solved in the low energy limit. This calculation shows that the particular choice made for the parameter $\Lambda$ in the HFB equations permits to implicitly sums the ladder diagram in the diagonal self-energy at a collisional energy consistent with the evaluation of the off-diagonal self energy.

\section{Interaction potential of vanishing range}

\subsection{Contact condition}

This paper deals with dilute gases of neutral bosons moving in a $D$-dimensional space (only the cases $D=2$ or $D=3$ are considered)  where interactions between particles can be described by a short range pairwise potential of radius denoted by ${b_D}$.  At sufficiently low density (denoted $n$) and  temperature,  \emph{i.e.} ${n b_D^3\ll 1}$ and ${\lambda_{DB} \gg b_D}$ where ${\lambda_{DB}}$ is the de Broglie wavelength, the transition matrix in two-body collisions of particles in the gas can be approximated by its lowest order in the binary collisional energy (denoted $E$):
\begin{equation}
T_D(E+i0^{+}) = \frac{\Omega_D \hbar^2}{2\mu} \times 
\left\{
\begin{array}{ll}
\frac{\displaystyle a_3}{\displaystyle(1 +  i a_3 k_0)};  & (D=3)\\
\\
\frac{\displaystyle -1}{\displaystyle\ln(-i a_2 k_0 e^\gamma/2)}; &  (D=2),\\
\end{array}
\right.
\label{eq:TD}
\end{equation}
where $k_0$ is the relative momentum (${E=\frac{\hbar^2k_0^2}{2\mu}}$), $\mu$ is the reduced mass,  ${a_D}$ is the $D$-dimensional scattering length and ${\Omega_D}$ is the full $D$-dimensional space angle (${\Omega_3=4\pi}$ and  ${\Omega_2=2\pi}$). In Eq.~\eqref{eq:TD} in the two-dimensional case, ${\gamma}$ is the Euler's constant, and the 2D scattering length $a_2$ is by definition a positive parameter. The zero range potential approach is a way to exactly recover these expressions of the transition matrix. It can be formulated for a system composed of $N$-particles as follows. The many-body wave-function in the configuration space is singular at the contact (${\mathbf r_i = \mathbf r_j}$) of each pair of interacting particles labeled by ${(ij)}$ while out of these singularities it solves the Schr\"{o}dinger's equation without pairwise interacting potential. For each interacting pair ${(ij)}$ as the relative coordinates (denoted ${\mathbf r_{ij}}$) tends to zero the singular behavior is parameterized by the scattering length with:
\begin{equation}
\Psi(\mathbf r_1, \dots \mathbf r_N) =  A_\Psi^{i\leftrightharpoons j} \times
\left\{ 
\begin{array}{ll}
\displaystyle  \left( \frac{1}{a_3} - \frac{1}{r_{ij}}\right) + O(r_{ij}); & {(D =3)} \\
\\
\displaystyle \ln\left(\frac{r_{ij}}{a_2}\right)  + O(r_{ij}); & {(D =2).} \\ 
\end{array}
\right. 
\label{eq:contact_D}
\end{equation}
In Eq.~\eqref{eq:contact_D} the limit is taken for fixed values of ${\mathbf r_k}$ (${k\ne i,j}$) and of the center of mass of the pair ${(ij)}$ and the strength's singularity denoted by ${A_\Psi^{i\leftrightharpoons j}}$ is a function of these coordinates. In this formulation often called Bethe Peierls model in the literature, the pairwise potential is thus replaced by asymptotic conditions on the wavefunction. While its simplicity is appealing, the Bethe Peierls model is not adapted to the standard techniques used in many-body theories.

\subsection{Pseudo-potential}

Another way to implement the zero-range potential approach is to use the $\Lambda$-potential introduced in Ref.~\cite{Ols01}. In this formalism, the true finite range pairwise potential is replaced by a pseudo-potential which is in fact a way to perform implicitly a subtractive regularizing scheme achieved at the negative energy ${E_\Lambda=-\frac{\hbar^2\Lambda^2}{2\mu}}$ where $\Lambda$ is an arbitrary positive parameter \cite{Pri11}. It can be introduced by its action on a two-body wavefunction $\psi$ as:
\begin{equation}
\langle \mathbf r_1, \mathbf r_2 | V_\Lambda | \psi \rangle  = T_D(E_\Lambda) {\mathcal R}_\Lambda\left[\psi\right] \delta(\mathbf r_{12}) ,
\label{eq:Lambda_potential}
\end{equation}
where ${T_D(E_\Lambda)}$ is evaluated at negative energy in Eq.~\eqref{eq:TD} by using the usual analytical continuation ${k_0=i\Lambda}$ and ${\mathcal R}_\Lambda\left[\,.\,\right]$ is a regularizing operator:
\begin{equation}
{\mathcal R}_\Lambda \left[ \, \cdot \, \right]= \lim_{r_{12} \to 0} 
\left\{
\begin{array}{ll}
\left( \partial_{r_{12}} + \Lambda \right) \left[ r_{12} \, \cdot \, \right]; &(D=3) \\
\\
\left[ 1- \ln \left(\frac{e^\gamma}{2} \Lambda r_{12}\right) r_{12} \partial_{r_{12}} \right] \left[ \, \cdot \, \right] & (D=2). \\
\end{array}
\right. 
\label{eq:regular}
\end{equation}
In Eq.~\eqref{eq:regular}, the subtraction cannot be performed at ${\Lambda=0}$ in 2D while in 3D the $\Lambda$ potential coincides with the Fermi-Huang pseudo-potential for ${\Lambda=0}$.

In Ref.~\cite{Pri11}, it has been shown that in two- and few-body problems it can be fruitful to relax the strictly zero range limit by replacing the $\delta$-distribution in Eq.~\eqref{eq:Lambda_potential} with a family of functions ${\delta_\epsilon}$ of finite support of range $\epsilon$ while keeping the subtractive structure of the $\Lambda$ potential and to perform the zero range limit ${\epsilon \to 0}$ in a second step. It appears convenient to choose a family of Gaussian functions such that ${\lim_{\epsilon  \to 0} \langle {\mathbf r} | \delta_\epsilon \rangle =  \delta({\mathbf r})}$ which have the interesting property in the momentum representation to have an expression independent of the dimension $D$:
\begin{equation}
\langle {\mathbf k}| \delta_\epsilon \rangle = \chi_\epsilon(k) = \exp \left(-\frac{k^2 \epsilon^2}{4} \right) .
\end{equation}
Using this family of functions, the ${\Lambda-\epsilon}$ potential can be written in the momentum representation as:
\begin{multline}
\langle \mathbf k_1,\mathbf k_2 | V^\Lambda_\epsilon | \mathbf k_3,\mathbf k_4 \rangle  \left[ \, \cdot \, \right] = (2\pi)^D \delta(\mathbf K_{12} -\mathbf K_{34})\\ 
T_D(E_\Lambda) \chi_\epsilon(\mathbf k_{12}) 
\lim_{\epsilon\to 0} r_\epsilon^\Lambda \left[ \chi_\epsilon(\mathbf k_{34}) \, \cdot \, \right] .
\label{eq:VLambda_epsilon_k}
\end{multline}
In Eq.~\eqref{eq:VLambda_epsilon_k} $\mathbf k_i$ is the momentum of the particle $i$, $\mathbf K_{ij}$ (respectively $\mathbf k_{ij}$) is the total (respectively the relative) momentum of the pair $(ij)$ and ${r^\Lambda_\epsilon}$ is a regularizing operator: 
\begin{equation}
r^\Lambda_\epsilon \left[ \, \cdot \, \right]=
\left\{ 
\begin{array}{ll}
\displaystyle \left[ \left(\partial_\epsilon +\sqrt{\frac{\pi}{2}} \Lambda \right) \epsilon  \, \cdot  \, \right]; & {(D=3)} \\
\\
\displaystyle \left[ \left( 1 + \frac{\epsilon}{2} \ln ( e^\gamma \Lambda^2 \epsilon^2/2 ) \partial_\epsilon \right)  \, \cdot  \, \right];   & {(D=2).} \\
\end{array}
\right. 
\label{eq:r_lambda_epsilon_expr}
\end{equation}
One may recognize in Eq.~\eqref{eq:VLambda_epsilon_k} a form similar to a separable potential. This structure is preserved  in the configuration space:
\begin{multline}
\langle \mathbf r_1,\mathbf r_2 |V^\Lambda_\epsilon | \mathbf r_3,\mathbf r_4 \rangle  \left[ \, \cdot \, \right]= \delta(\mathbf R_{12} - \mathbf R_{34})
\\ T_D(E_\Lambda) \langle \mathbf r_{12} | \delta_\epsilon\rangle
\lim_{\epsilon\to 0} r_\epsilon^\Lambda \left[ \langle  \delta_\epsilon | \mathbf r_{34} \rangle \, \cdot \, \right] ,
\label{eq:VLambda_epsilon_r}
\end{multline}
where ${\mathbf R_{ij}}$ (respectively ${\mathbf r_{ij}}$) denotes the center of mass (respectively the relative) coordinates of the pair $(ij)$ and ${\langle {\mathbf r}| \delta_\epsilon \rangle = (2\pi\epsilon^2)^{-D/2} \exp \left(-r^2/\epsilon^2\right)}$.

It is worth pointing out that the way the zero range limit is performed in the ${\Lambda-\epsilon}$ potential in Eqs.~(\ref{eq:VLambda_epsilon_k},\ref{eq:VLambda_epsilon_r}) totally differs from the usual method of the renormalization of the coupling constant (which tends to zero in the limit of a large ultraviolet momentum cutoff) often used in the many-body problem, even if the resulting equations coincide when the zero-range limit is well defined. This issue is considered in the appendix \ref{ap:renormalization}.

\section{HFB formalism for bosons}

\subsection{Variational approach in the zero range limit}

The asymptotic conditions in Eq.~\eqref{eq:contact_D} define the domain of the Hilbert space in the zero-range  limit. In principle, in a variational approach one must preserve this structure by using an ansatz in the correct Hilbert space. However in the many-body problem, this constraint appears too stringent for being adopted. In this respect, the introduction of the ${\Lambda-\epsilon}$ potential greatly simplifies the implementation of the variational schemes: one performs the variation of the functional at finite $\epsilon$ avoiding thus all the difficulties linked to the singularities of Eq.~\eqref{eq:contact_D} and the zero range limit where ${\epsilon=0}$ is only achieved when one uses the equations resulting from the variational principle. In the present section, this techniques is applied to the Hartree Fock Bogoliubov approach.

\subsection{Variation of the functional}

Using the expressions of the ${\Lambda-\epsilon}$ potential in Eqs.~(\ref{eq:VLambda_epsilon_k},\ref{eq:VLambda_epsilon_r}) the HFB equations are obtained at finite $\epsilon$ along the same lines as in Ref.~\cite{Bla86}. This section gathers the major steps of this derivation.

The gas is composed of $N$ identical bosons of  mass $m$. The field operator denoted by $\hat{\psi}$ verifies the usual bosonic commutation rules: ${[\hat{\psi}(\mathbf r),\hat{\psi}^\dagger(\mathbf r^\prime)]=(2\pi)^D \delta(\mathbf r -\mathbf r^\prime)}$ and  ${[\hat{\psi}(\mathbf r),\hat{\psi}(\mathbf r^\prime) ]=0}$. The standard HFB approximation is a symmetry breaking approach where the atomic field ${\hat{\psi}}$ is split into a classical field ${\Phi}$ and a quantum fluctuation 
\begin{equation}
\hat{\phi} = \hat{\psi} - \Phi \quad \mbox{with,} \quad  \langle \hat{\phi} \rangle = 0 .
\end{equation}
The density operator is chosen as a Gaussian ansatz in terms of the field $\hat{\phi}$:
\begin{equation}
\hat{D}_\epsilon = \frac{\exp(-\hat{K}_\epsilon/k_{\rm B}T)}{Z_\epsilon} ,
\label{eq:density_operator}
\end{equation}
where ${Z_\epsilon}$ is the partition function and the variational Hamiltonian ${\hat{K}_\epsilon}$ is quadratic in terms of $\{\hat{\phi},\hat{\phi}^\dagger\}$:
\begin{multline}
\hat{K}_\epsilon =  \int d^D r_1 d^D r_2
 \, \big[ h_\epsilon(\mathbf r_1,\mathbf r_2) \hat{\phi}^{\dagger}(\mathbf r_1) \hat{\phi}(\mathbf r_2) \\
+ \frac{1}{2} \Delta_\epsilon(\mathbf r_1,\mathbf r_2) \hat{\phi}^\dagger(\mathbf r_1) \hat{\phi}^\dagger(\mathbf r_2)\\
+ \frac{1}{2} \Delta_\epsilon^*(\mathbf r_1,\mathbf r_2)  \hat{\phi}(\mathbf r_1) \hat{\phi}(\mathbf r_2)  
\big] . 
\label{eq:ansatz}
\end{multline}
This ansatz is an implicit function of the classical field ${\Phi}$. The choice of a Gaussian density operator permits to use Wick's theorem. Consequently all the averages of products of the operator ${\hat{\phi}(\mathbf r)}$ can be expressed in terms of the diagonal one-particle density matrix
\begin{equation}
\tilde{\rho}_\epsilon(\mathbf r_1,\mathbf r_2)=\langle \hat{\phi}^\dagger(\mathbf r_2)\hat{\phi}(\mathbf r_1)\rangle_\epsilon
\label{eq:rho_tilde}
\end{equation}
and of the off-diagonal one-particle density matrix
\begin{equation}
\tilde{\kappa}_\epsilon(\mathbf r_1,\mathbf r_2) =  \langle \hat{\phi}(\mathbf r_2)\hat{\phi}(\mathbf r_1) \rangle_\epsilon .
\label{eq:kappa_tilde}
\end{equation}
In Eqs.~(\ref{eq:density_operator}-\ref{eq:kappa_tilde})  and in what follows, quantities or averages where an index $\epsilon$ is added means that they are considered for a finite value of $\epsilon$ while the index $\epsilon$ is dropped when the zero range limit ${(\epsilon \to 0)}$ is achieved. For example, this is the case for the depletion denoted by ${\tilde{n}(\mathbf R)=\tilde{\rho}(\mathbf R,\mathbf R)}$ and for the total particle density
\begin{equation}
n(\mathbf R) = |\Phi(\mathbf R)|^2 + \tilde{n}(\mathbf R) .
\end{equation}
In the subsequent equations the pair wavefunction is used:
\begin{equation}
\kappa_\epsilon(\mathbf r_1,\mathbf r_2) =  \langle \hat{\psi}(\mathbf r_2) \hat{\psi}(\mathbf r_1) \rangle_\epsilon 
\end{equation}
and also the $\Lambda$-regularized anomalous averages
\begin{eqnarray}
&&\tilde{\kappa}^\Lambda(\mathbf R_{12}) = \lim_{\epsilon\to 0} r_\epsilon^\Lambda \int d^D r_{12} 
\langle \mathbf r_{12} | \delta_\epsilon \rangle \tilde{\kappa}_\epsilon(\mathbf r_1,\mathbf r_2) \\
\label{eq:kappa_Lambda_tilde}
&&\kappa^\Lambda(\mathbf R_{12}) = \Phi(\mathbf R_{12})^2 + \tilde{\kappa}^\Lambda(\mathbf R_{12}) .
\end{eqnarray}
In Eq.~\eqref{eq:kappa_Lambda_tilde} ${\tilde{\kappa}^\Lambda}$ can be also evaluated from Eq.~\eqref{eq:regular} by ${\tilde{\kappa}^\Lambda=\mathcal R_\Lambda \left[\tilde{\kappa} \right]}$. The variational principle applied to the Grand-potential ${E_\epsilon-TS_\epsilon-\tilde{\mu} N_\epsilon}$ where $E_\epsilon$ is the total energy, $N_\epsilon$ is the number of particles and ${S_\epsilon=-k_{\rm B} \langle \ln(\hat{D_\epsilon})\rangle_\epsilon}$ is the entropy gives \cite{Bla86}:
\begin{multline}
h_\epsilon(\mathbf r_1,\mathbf r_2) = -\frac{\hbar^2}{2m} \left(\Delta_{\mathbf r}\delta \right)(\mathbf r_{12})-\tilde{\mu}\delta(\mathbf r_{12})+ V_{\rm ext}(\mathbf R_{12})\\
+  2 \int d^D r_3 d^D r_4 \langle \mathbf r_1,\mathbf r_3 | V^\Lambda_\epsilon | \mathbf r_2,\mathbf r_4 \rangle \rho_\epsilon(\mathbf r_4, \mathbf r_3)
\end{multline}
and
\begin{equation}
\Delta_\epsilon(\mathbf r_1,\mathbf r_2) =  \int d^D r_3 d^D r_4
\langle \mathbf r_1,\mathbf r_2 | V^\Lambda_\epsilon | \mathbf r_3,\mathbf r_4 \rangle \kappa_\epsilon(\mathbf r_3, \mathbf r_4) .
\end{equation}
After a straightforward integration one obtains:
\begin{multline}
h_\epsilon(\mathbf r_1,\mathbf r_2) = -\frac{\hbar^2}{2m} \left(\Delta \delta \right)(\mathbf r_{12}) + \left[V_{\rm ext}(\mathbf R_{12})-\tilde{\mu}\right] \delta(\mathbf r_{12}) \\ 
+ \hbar \Sigma_{11}(\mathbf R_{12}) \langle \mathbf r_{12} | \delta_\epsilon \rangle 
\end{multline}
and
\begin{equation}
\Delta_\epsilon(\mathbf r_1,\mathbf r_2) = \hbar \Sigma_{12}(\mathbf R_{12}) \langle \mathbf r_{12} | \delta_\epsilon \rangle ,
\end{equation}
where the diagonal and off-diagonal self-energies are defined by:
\begin{equation}
\left\{
\begin{array}{l}
\hbar \Sigma_{11}(\mathbf R) = 2 T_D(E_\Lambda) n(\mathbf R) , \\
\hbar \Sigma_{12}(\mathbf R) =  T_D(E_\Lambda) \kappa^\Lambda(\mathbf R)  .
\end{array}
\right.
\label{eq:Sigma_HFB}
\end{equation}
The  equation verified by the condensate field is also obtained easily at finite ${\epsilon}$ by minimization of the Grand potential. In presence of an external potential ${V_{\rm ext}(\mathbf r)}$ acting on each particle of the gas, in the zero range limit it reads:
\begin{multline}
(H_{\mathbf r}^{\rm 1p} - \tilde{\mu} ) \Phi(\mathbf r) + T_D(E_\Lambda) \left[ 2 \tilde{n}(\mathbf r) + |\Phi(\mathbf r)|^2 \right] \Phi(\mathbf r) \\
+ T_D(E_\Lambda) \tilde \kappa^\Lambda(\mathbf r) \Phi^*(\mathbf r) = 0 .
\label{eq:GP_HFB}
\end{multline}
In Eq.~\eqref{eq:GP_HFB}, the one-particle Hamiltonian ${H_{\mathbf r}^{\rm 1p}}$ is defined by:
\begin{equation}
H_{\mathbf r}^{\rm 1p}=-\frac{\hbar^2}{2m} \Delta_{\mathbf r} + V_{\rm ext}(\mathbf r) .
\end{equation}
The variational ansatz $\hat{K}_\epsilon$ is diagonalized by using a special Bogoliubov transformation on the particle field:
\begin{equation}
\hat{\phi}(\mathbf r) = \sum_n \left[ \hat{b}_n u_n(\mathbf r) + \hat{b}_n^\dagger v_n^*(\mathbf r) \right] ,
\label{eq:transfo_Bogo}
\end{equation}
where the bosonic operator $\hat{b}_n$ (resp. $\hat{b}_n^\dagger$) destroys (resp. creates) a quasi-particle of quantum number $n$. The modal amplitudes $\{u_n,v_n\}$ in Eq.~\eqref{eq:transfo_Bogo} are the eigenvectors of the linear system:
\begin{eqnarray} 
&&\left(H^{\rm 1p} - \tilde{\mu} + \hbar \Sigma_{11}  \right) u_n + \hbar \Sigma_{12} v_n = \hbar \omega_n u_n \nonumber\\
&&\left(H^{\rm 1p} - \tilde{\mu} + \hbar \Sigma_{11} \right) v_n^* + \hbar \Sigma_{12}^* u_n^* = - \hbar \omega_n v_n^* .
\label{eq:diff_modes}
\end{eqnarray}
In Eq.~\eqref{eq:diff_modes} the coordinates ${(\mathbf r)}$ have been omitted for conciseness. In this basis the variational ansatz can be written as:
\begin{equation}
\hat{K}_\epsilon= \sum_n  \hbar \omega_n b_n^\dagger b_n .
\end{equation}
From the commutation relations of the operators ${(\hat{b}_n,\hat{b}_n^\dagger,\hat{\phi},\hat{\phi}^\dagger)}$, one can deduce the closure relations 
\begin{eqnarray}
& &\sum_n \left[ u_n(\mathbf r_1)  u_n^*(\mathbf r_2) -  v_n^*(\mathbf r_1) v_n(\mathbf r_2) \right] =  \delta(\mathbf r_{12})  \label{eq:fermeture1}
\\
& &\sum_n \left[ v_n(\mathbf r_1)  u_n^*(\mathbf r_2) -  u_n^*(\mathbf r_1) v_n(\mathbf r_2) \right] = 0 .
\label{eq:fermeture2}
\end{eqnarray}
The two fields $\tilde{\rho}$ and $\tilde{\kappa}$ can be expressed in terms of the mode amplitudes at thermal equilibrium~:
\begin{eqnarray}
\tilde{\rho}(\mathbf r_1,\mathbf r_2) &=& \sum_n \big[ v_n^*(\mathbf r_1) v_n(\mathbf r_2) \left(1+f_n\right)+\nonumber \\
 && \qquad \qquad u_n(\mathbf r_1) u_n^*(\mathbf r_2) f_n
\big] \label{eq:rho_modal} \\
\tilde{\kappa}(\mathbf r_1,\mathbf r_2)&=& \sum_n \left(\frac{1}{2}+f_n \right) \big[  v^*_n(\mathbf r_1) u_n(\mathbf r_2) + \nonumber\\
&&\qquad \qquad  v^*_n(\mathbf r_2) u_n(\mathbf r_1) \big] 
\label{eq:kappa_modal},
\end{eqnarray}
where ${f_n=\langle b^\dagger_n b_n \rangle}$ is the Bose occupation factor:
\begin{equation}
f_n=\frac{1}{\exp(\frac{\hbar \omega_n}{k_B T})-1} .
\end{equation}
In the zero range limit, the one-particle density ${\tilde{\rho}(\mathbf r_1,\mathbf r_2)}$ is regular at $r_{12}=0$ and is equal to the depletion, while the pair wavefunction ${\kappa(\mathbf r_1,\mathbf r_2)}$ verifies the Bethe-Peierls contact condition in Eq.~(\ref{eq:contact_D}) as the relative coordinates ${r_{12}}$ tend to zero. The simplest way to show this property is to  derive the equation verified by $\tilde{\kappa}(\mathbf r_1,\mathbf r_2)$. Using Eqs.~(\ref{eq:diff_modes},\ref{eq:fermeture1}-\ref{eq:kappa_modal}) one obtains:
\begin{multline}
\left[ H_{\mathbf r_1}^{\rm 1p} + H_{\mathbf r_2}^{\rm 1p} \right] \kappa( \mathbf r_1, \mathbf r_2)
 + \langle \mathbf r_1, \mathbf r_2 | V^\Lambda | \kappa \rangle =\\ 
 2\tilde{\mu} \tilde{\kappa}( \mathbf r_1, \mathbf r_2)
-\left[ \hbar \Sigma_{11}(\mathbf r_1) + \hbar \Sigma_{11}(\mathbf r_2) \right] \tilde{\kappa}(\mathbf r_1, \mathbf r_2 )\\
- \hbar \Sigma_{12}(\mathbf r_1) \tilde{\rho}(\mathbf r_2,\mathbf r_1) - \hbar
\Sigma_{12}(\mathbf r_2) \tilde{\rho}(\mathbf r_1,\mathbf r_2)  .
\label{eq:eq_pair}
\end{multline}
In the right hand side of Eq.~\eqref{eq:eq_pair} there is no $\delta$-distribution while the $\Lambda$ potential acts on the pair wavefunction in the left hand side of this equation. Hence, the pair wavefunction  verifies the contact condition Eq.~(\ref{eq:contact_D})  and using Eq.~\eqref{eq:regular} one founds the strength's singularity in terms of the anomalous average:
\begin{equation}
A_\kappa(\mathbf R) = \frac{2\mu}{\hbar^2 \Omega_D} T_D(E_\Lambda) \kappa^{\Lambda}(\mathbf R) \quad \forall \Lambda >0 . 
\end{equation}
The off-diagonal self-energy evaluated from Eq.~\eqref{eq:Sigma_HFB} is then proportional to $A_\kappa$:
\begin{equation}
\hbar \Sigma_{12}(\mathbf R) =\frac{\hbar^2 \Omega_D}{2\mu} A_\kappa(\mathbf R)
\label{eq:Sigma12_invariance}
\end{equation}
and is thus not an explicit function of the parameter $\Lambda$.

\subsection{A judicious choice for ${\Lambda(\mathbf R)}$}

In an exact approach, the description of the system does not depend on a particular choice made for the free parameter $\Lambda$ [which can be also a function of ${\mathbf R_{12}}$ in Eq.~\eqref{eq:VLambda_epsilon_r}]. However, the expression of the diagonal self-energy in Eq.~\eqref{eq:Sigma_HFB} shows that HFB equations are $\Lambda$-dependent. Interestingly, this makes possible to solve the so-called gap problem in HFB (or Gaussian variational approach) which is not compatible with the Hugenholtz-Pines theorem \cite{Hug59} and was first shown to occur in Ref.~\cite{Gir59}. For this purpose, the natural choice for the free-parameter $\Lambda$ is given by imposing that the anomalous mode ${(u_0,v_0)=(\Phi,-\Phi)}$ is a zero energy solution of Eq.~\eqref{eq:diff_modes}. In an inhomogeneous systems, this leads to a value of $\Lambda$ which depends on the position. This specific function  ${\Lambda(\mathbf R)}$  is denoted by ${\Lambda^\star(\mathbf R)}$ and is obtained by solving the equation:
\begin{equation}
\tilde{\kappa}^{\Lambda{^\star}}(\mathbf R) =0  \quad \forall \ \mathbf R.
\label{eq:Lambda_star}
\end{equation}
Using the $\Lambda$ freedom of the off-diagonal self energy, one finds 
\begin{equation} 
T_D(E_{\Lambda^\star})(\mathbf R) = T_D(E_\Lambda)(\mathbf R) \left[ 1 + \frac{\tilde{\kappa}^{\Lambda}(\mathbf R)}{\Phi^2(\mathbf R)} \right] .
\label{eq:g_Lambda_star}
\end{equation}
In the 3D case Eq.~\eqref{eq:g_Lambda_star} gives~\cite{Ols01}: 
\begin{equation}
{a \Lambda^\star(\mathbf R) = \frac{\tilde{\kappa}^{\Lambda=0}(\mathbf R)}{\tilde{\kappa}^{\Lambda=0}(\mathbf R) + \Phi^2(\mathbf R)}} ,
\label{eq:Lambda_star_3D}
\end{equation}
while in the 2D case there doesn't exist such a general expression.

\subsection{Homogeneous gas}

In what follows equations are used for the homogeneous gas with the choice ${\Lambda=\Lambda^\star}$, assuming thus that Eq.~\eqref{eq:Lambda_star} is satisfied. The thermodynamic limit is also supposed to be achieved. The bosonic field operator is expanded on the plane wave basis:
\begin{equation}
\hat{\psi}(\mathbf r) = c_0 + \int \frac{d^Dk}{(2\pi)^D} c_{\mathbf k} e^{i \mathbf k \cdot \mathbf r} ,
\end{equation}
where ${\Phi= \langle c_0 \rangle}$ and ${c_{\mathbf k}}$ annihilates a plane wave of momentum ${\mathbf k}$. The operators ${c_{\mathbf k}}, {c_{\mathbf k^\prime}^\dagger}$  verify the usual commutation rules:
${[c_{\mathbf k} ,c_{\mathbf k^\prime}^\dagger] =(2\pi)^D \delta(\mathbf k - \mathbf k^\prime)}$
and  ${[c_{\mathbf k},c_{\mathbf k^\prime}] = 0}$.  The special Bogoliubov transformation takes the form: 
\begin{equation}
\hat{\phi}(\mathbf r) = \int \frac{d^3k}{(2\pi)^3} \left( b_{\mathbf k} u_{\mathbf k} e^{i \mathbf k \cdot \mathbf r} + b_{\mathbf k}^\dagger v_{\mathbf k}^* e^{-i \mathbf k \cdot \mathbf r}  \right) ,
\end{equation}
where the operators ${(b_{\mathbf k} ,b_{\mathbf k}^\dagger)}$ respectively create and destroy a quasi-particle of momentum $\mathbf k$ and verify also the bosonic commutation rules ${[b_{\mathbf k},b_{\mathbf k^\prime}^\dagger ]  = (2\pi)^D \delta(\mathbf k-\mathbf k^\prime)}$, ${[b_{\mathbf k},b_{\mathbf k^\prime} ]  = 0}$ and the modal amplitudes verify
\begin{equation}
|u_{\mathbf k}|^2 - |v_{\mathbf k}|^2 = 1 .
\end{equation}
Equations~(\ref{eq:GP_HFB},\ref{eq:Lambda_star}) give the expression of the chemical potential for the homogeneous gas in terms of the condensate density and of the depletion:
\begin{equation}
{\tilde{\mu}= T_D(E_{\Lambda^\star}) ( \Phi^2 + 2 \tilde n)} .
\label{eq:mu_tilde}
\end{equation}
Hence using Eqs.~(\ref{eq:Sigma_HFB},\ref{eq:mu_tilde}) one obtains for ${\Lambda=\Lambda^\star}$ the relation ${\tilde{\mu} =  \hbar \Sigma_{11} - \hbar \Sigma_{12}}$, which is the Hugenholtz Pines theorem \cite{Hug59}. The solutions of Eq.~\eqref{eq:diff_modes} are chosen real:
\begin{eqnarray}
&&u_{\mathbf k} = \frac{1}{\sqrt{2}} \left[ \frac{\displaystyle \frac{\hbar^2 k^2}{2m} + T_D(E_{\Lambda^\star})\Phi^2}{\hbar \omega_k} + 1 \right]^{\frac{1}{2}} 
\nonumber\\
&&v_{\mathbf k} = -\frac{1}{\sqrt{2}} \left[ \frac{\displaystyle \frac{\hbar^2 k^2}{2m} + T_D(E_{\Lambda^\star})\Phi^2}{\hbar \omega_k} - 1 \right]^{\frac{1}{2}} \label{eq:modes}\\
&&\hbar \omega_{\mathbf k} = \left( \frac{\hbar^2 k^2}{2m} \right)^{\frac{1}{2}} \left[ \frac{\hbar^2 k^2}{2m} + 2 T_D(E_{\Lambda^\star}) \Phi^2 \right]^{\frac{1}{2}} .
\nonumber
\end{eqnarray}
At zero temperature the depletion is given by 
\begin{equation}
\tilde{n}= \int \frac{d^D k}{(2\pi)^D} |v_{\mathbf k}|^2 
\end{equation}
and in the zero-range limit, the off-diagonal one-particle density matrix is:
\begin{equation}
\tilde{\kappa}(\mathbf r_1,\mathbf r_2) = \int \frac{d^Dk}{(2\pi)^D} \frac{-\hbar\Sigma_{12}}{2\hbar \omega_{\mathbf k}} \exp(i\mathbf k \cdot \mathbf r_{12}) .
\end{equation}
HFB solutions are found for a given value of the particle density:
\begin{eqnarray}
n = \Phi^2 + \int \frac{d^Dk}{(2\pi)^D}\, v_{\mathbf k}^2  .
\label{eq:norm} 
\end{eqnarray}

\subsubsection{3D Bose gas}

The 3D homogeneous case has been studied in Ref.~\cite{Ols01}. At zero temperature, Eqs.~(\ref{eq:Lambda_star_3D},\ref{eq:modes},\ref{eq:norm}) together with the following expression of the regular part of the pairing field [obtained for example by using Eq.~\eqref{eq:r_espilon_subtractive}]:
\begin{equation}
\tilde \kappa^{\Lambda=0}= \hbar \Sigma_{12}  \int \frac{d^3k}{(2\pi)^3}\, \left( \frac{m}{\hbar^2 k^2}-\frac{1}{2\hbar\omega_{\mathbf k}} \right) 
\end{equation}
lead to 
\begin{equation}
(n - \Phi^2)a^3 = \frac{8}{3 \sqrt{\pi}} \left[ a^3(3n-2\Phi^2) \right]^{3 \over 2} . 
\label{eq:n_Phi}
\end{equation}
Hence, for each value of the atomic density, the condensate fraction is found by solving equation (\ref{eq:n_Phi}). In the dilute regime (${na^3\ll1}$) one has  ${n\simeq\Phi^2}$, ${T_3(E_{\Lambda^\star}) \simeq T_3(0)}$ and as expected the variational approach coincides with the usual perturbative Bogoliubov theory~\cite{Ols01}. The transition matrix ${T_3(E_{\Lambda^\star})}$ is an increasing function of the density and diverges at the critical value ${\displaystyle n_c=\frac{\pi}{192 a^3}}$ where ${\Lambda^\star=1/a}$ and ${\Phi=0}$. In this regime  the mean field approach breaks-down, ${a\Lambda^\star}$ is close to unity and the probability density for a quasi-particle to have a wavenumber $k$ [\emph{i.e.}  ${v_{\mathbf k}^2 k^2/(2\pi^2\tilde n)}$] is peaked near ${k=1/a}$: collisions between non condensed atoms are not well described in this formalism. Clearly, inelastic processes and correlations not included in the HFB approximation must be taken into account to study this regime. 

\subsubsection{2D Bose gas}

In the 2D  homogeneous case, the standard HFB formalism for bosons can be used only at zero  temperature where a true condensate exists. This situation was studied in Ref.~\cite{Pri04b} where it was shown that the present HFB equations permit to obtain an EOS which appears recently to be very good when compared to exact perturbative results beyond the  Bogoliubov approximation. In what follows, this EOS is derived in a parametric form. For this purpose, one first calculates the asymptotic behavior of the off-diagonal one-particle density matrix as ${r_{12}}$ tends to zero:
\begin{equation}
\tilde{\kappa}(\mathbf r_1,\mathbf r_2) = \frac{m \Sigma_{12}}{2\pi\hbar} \ln \left[\frac{e^\gamma }{2\hbar} \sqrt{mT_2(E_{\Lambda^\star})} |\Phi| \times r_{12} \right]
+ O (r_{12}) 
\end{equation}
which provides the ${\Lambda}$-regularized part:
\begin{equation}
\tilde{\kappa}^{\Lambda}= \frac{m \Sigma_{12}}{4\pi\hbar} \ln \left[ \frac{m T_2(E_{\Lambda^\star}) |\Phi|^2}{\hbar^2\Lambda^2}\right] .
\label{eq:tilde_kappa_Lambda}
\end{equation}
The consistent choice for ${\Lambda^\star}$ obtained from Eq.~\eqref{eq:Lambda_star} is thus
\begin{equation}
\Lambda^\star = \sqrt{\frac{-2\pi |\Phi|^2 }{\ln\left(e^\gamma \Lambda^\star a_2/2 \right)}} .
\end{equation}
The depletion is given by
\begin{equation}
\tilde{n}= \frac{m T_2(E_{\Lambda^\star})|\Phi|^2}{4\pi \hbar^2} = \frac{\Lambda^\star\,^2}{4\pi} .
\end{equation}
Finally using Eq.~\eqref{eq:mu_tilde}, the constraint ${\Lambda=\Lambda^\star}$ imposed by the Hugenholtz Pines theorem provides the EOS
\begin{equation}
n = \frac{t^2 \left[1-2 \ln \left( \frac{e^\gamma t}{2} \right) \right]}{4\pi a_2^2}  \ ; \ 
\tilde{\mu} = \frac{\hbar^2t^2 \left[ 1 - \frac{1}{\ln \left( \frac{e^\gamma t}{2} \right)} \right]}{m a_2^2} ,
\label{eq:EOS_2D_1}
\end{equation}
defined through the dimensionless parameter ${t=a_2 \Lambda^\star}$ (${0<t<2e^{-\gamma}}$). The energy per particle is obtained from ${E/N=\int_0^n \tilde{\mu} dn}$
\begin{equation}
\frac{E}{N}=\frac{\hbar^2t^2}{ma_2^2} \times \frac{\frac{5}{4}-\ln \left( \frac{e^\gamma t}{2} \right) }{1-2\ln \left( \frac{e^\gamma t}{2} \right)} .
\label{eq:E0S_2D_2}
\end{equation}
The low areal density regime corresponds to the limit of vanishing values of the dimensionless parameter ${t \to 0^+}$ and in this regime one recovers the result of Popov~\cite{Pop72}:
\begin{equation}
n  = \frac{m \tilde{\mu}}{4\pi \hbar^2}\ln\left[ \frac{4 \hbar^2}{e^{(2\gamma+1)}m \tilde{\mu} a_2^2} \right] \quad \mbox{for} \ n a_2^2 \ll 1 ,
\label{eq:popov}
\end{equation}
which is also the prediction of the Bogoliubov theory \cite{Mor03}. For finite areal densities, Eq.~\eqref{eq:EOS_2D_1} gives an estimate compatible with Monte-Carlo simulations \cite{Pil05,Ast09}, the perturbative expansion one order beyond the Bogoliubov theory \cite{Mor09} and another approximate method \cite{Che01}.

\subsubsection{An implicit ladder summation}

In terms of Feynman diagrams, the diagonal self-energy  ${\hbar \Sigma_{11}}$ in Eq.~\eqref{eq:Sigma_HFB} is treated at the one-loop level \cite{Fet03}. That is why it explicitly depends on $\Lambda$ [unlike the off-diagonal self-energy as shown in Eq.~\eqref{eq:Sigma12_invariance}]. This section shows that the choice ${\Lambda=\Lambda^\star}$ permits to implicitly perform the summation of the ladder diagrams enabling thus to evaluate the $T$-matrix appearing in ${\hbar \Sigma_{11}}$ at the correct energy.

In the standard many-body technique, the collision between two quasi-particles is described by the effective potential $\Gamma$, which takes into accounts the effect of the medium \cite{Fet03}:
\begin{equation}
\langle \underline{k}_1,\underline{k}_2 |\Gamma|\underline{k}_3,\underline{k}_4 \rangle ,
\end{equation}
where  ${\underline{k}=(\omega,\mathbf k)}$ denotes a $D+1$ dimensional vector. In subsequent equations, the bare vertex for the ${\Lambda-\epsilon}$ potential is denoted by:
\begin{equation}
\langle \underline{k}_1,\underline{k}_2 | \Gamma^b_\epsilon | \underline{k}_3,\underline{k}_4\rangle = \langle \mathbf k_1,\mathbf k_2 | V^\Lambda_\epsilon | \mathbf k_3,\mathbf k_4 \rangle .
\label{eq:Gamma_0_epsilon}
\end{equation}
The ladder summation is obtained through the ladder approximation of the Bethe Salpeter equation \cite{Fet03}:
\begin{multline}
\langle \underline{k}_1,\underline{k}_2 |\Gamma_\epsilon |\underline{k}_3,\underline{k}_4 \rangle 
= \langle \underline{k}_1,\underline{k}_2 | \Gamma^b_\epsilon | \underline{k}_3,\underline{k}_4\rangle  + \frac{i}{\hbar}
\int \,  \frac{d^{D+1} q}{(2\pi)^{D+1}} \\
\times \langle \underline{k}_1,\underline{k}_2 | \Gamma^b_\epsilon | \underline{k}_1- \underline{q},\underline{k}_2+\underline{q} \rangle G(\underline{k}_1- \underline{q}) \\ 
\times   G(\underline{k}_2+\underline{q})
\langle  \underline{k}_1-\underline{q},\underline{k}_2+\underline{q}  | \Gamma_\epsilon | \underline{k}_3,\underline{k}_4\rangle ,
\label{eq:Bethe-Salpeter}
\end{multline}
where ${\underline{q}=(q_0,\mathbf q)}$. In what follows, this equation is solved at zero temperature by using the one-body Green's function consistent with the HFB approximation:
\begin{equation}
G(\underline{k}) = \frac{u_{\mathbf k}^2}{\omega-\omega_{\mathbf k}+i 0^+} - \frac{v_{\mathbf k}^2}{\omega+\omega_{\mathbf k} -i 0^+} .
\label{eq:green}
\end{equation}
The structure of Eq.~\eqref{eq:Bethe-Salpeter} together with the definition of $\Gamma^b_\epsilon$ in Eqs.~(\ref{eq:VLambda_epsilon_k},\ref{eq:Gamma_0_epsilon}), give the form of the dressed vertex:
\begin{multline}
\langle \underline{k}_1,\underline{k}_2 |\Gamma_\epsilon |\underline{k}_3,\underline{k}_4 \rangle \left[ \, \cdot \, \right] = (2\pi)^D \delta(\mathbf K_{12} -\mathbf K_{34})\\ 
g_{\rm eff} \chi_\epsilon(\mathbf k_{12}) 
\lim_{\epsilon\to 0} r_\epsilon^\Lambda \left[ \chi_\epsilon(\mathbf k_{34}) \, \cdot \, \right]
\label{eq:ansatz_Gamma}
\end{multline}
and one wishes to compute the strength ${g_{\rm eff}}$. Integration over $q_0$ gives:
\begin{multline}
\frac{i}{\hbar} \int \frac{dq_0}{2\pi} \, G(\underline{k}_1- \underline{q}) G(\underline{k}_2+\underline{q}) =\\
\frac{  u_{\mathbf k_1- \mathbf q}^2 u_{\mathbf k_2 + \mathbf q}^2}
{-\hbar \Omega+ \hbar \omega_{\mathbf k_1- \mathbf q}+\hbar\omega_{\mathbf k_2 + \mathbf q}} 
+ \frac{  v_{\mathbf k_1- \mathbf q}^2 v_{\mathbf k_2 + \mathbf q}^2}{\hbar\Omega+\hbar\omega_{\mathbf k_1 - \mathbf q} + \hbar \omega_{\mathbf k_2 + \mathbf q}} ,
\end{multline}
where $\Omega=\omega_1+\omega_2$. In the limit of vanishing energy: ${k_1 \to  0}$ , ${k_2 \to  0}$ and ${\Omega\to0}$ one has:
\begin{equation}
g_{\rm eff}= \frac{T_D(E_\Lambda)}{1-T_D(E_\Lambda) J} ,
\label{eq:geff}
\end{equation}
where
\begin{equation}
J = - \lim_{\epsilon \to 0}r_\epsilon^\Lambda\left[ \int \frac{d^Dq}{(2\pi)^D} \chi_\epsilon(q)^2 \frac{u_{\mathbf q}^4+v_{\mathbf q}^4}{2 \hbar \omega_{\mathbf q}} \right].
\label{eq:regul_Ladder}
\end{equation}	
One then may use the identities:
\begin{equation}
 u_{\mathbf q}^4+v_{\mathbf q}^4 = 1 + 2 u_{\mathbf q}^2 v_{\mathbf q}^2  \quad \mbox{and} \quad u_{\mathbf q} v_{\mathbf q}=-\frac{\hbar \Sigma_{12}}{2\hbar\omega_{\mathbf q}} .
\end{equation}
The bit associated with ${u_{\mathbf q}^2 v_{\mathbf q}^2}$ in Eq.~\eqref{eq:regul_Ladder} gives an unphysical infrared divergence due to the fact that the effective potential is evaluated here in the strict zero energy limit of the scattering processes. In what follows this contribution on the value of $J$ is denoted by ${J_{\rm div}}$. A natural infrared cut-off for ${J_{\rm div}}$ consistent with the ladder approximation in the medium is given by the inverse inter-particle distance ${q_{\rm min} \sim n^{1/D}}$. Owing to the inequality ${q_{\rm min} \gg 1/\xi}$ where ${\xi=\sqrt{\hbar/(2m\Sigma_{12})}}$ is the healing length in the Bose gas one finds that ${T_D(E_\Lambda) J_{\rm div} \equiv O(q_{\rm min}^{-6} \xi^{-6})}$ can be neglected  in Eq.~\eqref{eq:geff}. The approximation ${u_{\mathbf q}^4+v_{\mathbf q}^4 \sim1}$ can thus be performed,   and one may recognize in Eq.~\eqref{eq:regul_Ladder} the expression of the $\Lambda$-regularized part of the off-diagonal one-particle density matrix at zero temperature considered for finite values of $\epsilon$:
\begin{equation}
 \tilde{\kappa}_\epsilon(\mathbf r_1,\mathbf r_2)= \int \frac{d^Dq}{(2\pi)^D} \chi_\epsilon(q) \frac{-\hbar\Sigma_{12}}{2 \hbar \omega_{\mathbf q}} \exp(i\mathbf q \cdot \mathbf r_{12}) .
\end{equation}
One thus obtains ${J=\tilde{\kappa}^\Lambda/(\hbar  \Sigma_{12})}$, and using Eqs.~(\ref{eq:g_Lambda_star},\ref{eq:geff})  the strength appearing in the effective potential is given by:
\begin{equation}
g_{\rm eff} = T_D(E_\Lambda) \frac{\Phi^2+\tilde{\kappa}^\Lambda}{\Phi^2} = T_D(E_{\Lambda^\star}) . 
\end{equation}
Assuming that the spectrum is gapless, this result is valid for all $\Lambda$. For ${\Lambda=\Lambda^\star}$ then  ${\Gamma_\epsilon=\Gamma^b_\epsilon}$ in Eq.~\eqref{eq:Bethe-Salpeter}, meaning that when used at the one-loop order the bare vertex ${\Gamma^b_\epsilon}$ incorporates the ladder summation for ${\Lambda=\Lambda^\star}$, which is the desired result. 

\section{Conclusions}

In this paper the variational HFB equations in the zero range potential approach have been obtained following the standard derivation of Ref.~\cite{Bla86}. This way, results  of Refs.~\cite{Ols01,Pri04b} where the $\Lambda$ potential was used are recovered in a simple manner. The EOS for the 2D Bose gas obtained in the HFB  approximation is expressed in a parametric form.  It is  explicitly shown that the specific choice made for the parameter $\Lambda$ permits to perform the ladder summation while keeping the variational scheme intact. These results have been obtained by using a representation of the $\Lambda$ potential where the strict zero range limit is relaxed in intermediary calculations. The natural continuation of this work is to use the ${\Lambda-\epsilon}$ potential vertex of Eq.~\eqref{eq:Gamma_0_epsilon} in standard techniques of many-body theory borrowed from quantum field theory, including the treatment of the fermionic gas. The $\Lambda$ freedom is a gift of the zero-range potential approach which permits to improve approximate treatments of the many-body problem. The best choice for the parameter $\Lambda$  in the general time dependent HFB formalism or also in the variational Balian V\'{e}n\'{e}roni approach \cite{Bla86} remains an open question.  Another  interesting issue is the possible use of a fully variational approach for the 2D Bose gas at finite temperature where the low energy classical modes of the field operator could be treated consistently in the variational scheme. 

\section*{Acknowledgments}

Yvan Castin, Maxim Olshanii and Peter Schuck are acknowledged for discussions. Laboratoire de Physique Th\'{e}orique de la Mati\`{e}re Condens\'{e}e is UMR 7600 of CNRS and its Cold Atoms group is associated with IFRAF.

\appendix

\section{Contact condition obtained from the renormalization of the coupling constant}
\label{ap:renormalization}

In the literature, numerous effective interactions are used for implementing the zero-range potential limit. As an example, the contact interaction on a discrete lattice where the coupling constant is adjusted to reproduce the low energy scattering properties is well adapted for numerical studies and leads to equations very similar to the one obtained using the contact conditions of Eq.~\eqref{eq:contact_D} \cite{Pri07b}. This appendix focuses on a standard techniques of the renormalization of the  vertex interaction in the 3D continuous space where the strength of the force vanishes in the zero range limit. It is shown below that in the zero-range limit of the effective interaction, one recovers in the two-body problem the contact condition of the Bethe Peierls model obtained from the ${\Lambda-\epsilon}$ potential in the momentum representation.

\subsection{Bethe Peierls contact condition in the momentum representation}
\label{ap:momentum_representation}

In Ref.~\cite{Pri11}, the operator ${r^\Lambda_\epsilon}$ was introduced without loss of generality through the identity:
\begin{equation}
\lim_{\epsilon \to 0} r^\Lambda_\epsilon \left[\langle \delta_\epsilon | \psi_\epsilon \rangle \right] 
= \lim_{\epsilon \to 0}  \left[ \langle \delta_\epsilon | \psi_\epsilon  -    
\frac{\hbar^2 \Omega_D}{2\mu}  A_\psi \phi_\epsilon^\Lambda \rangle \right] ,
\label{eq:r_espilon_subtractive}
\end{equation}
where ${|\psi_\epsilon\rangle}$ is a two-body state considered in the center of mass frame, ${A_\psi}$ is its strength's singularity in the limit where $\epsilon$ tends to zero, and ${|\phi_\epsilon^\Lambda \rangle}$ is a reference state equal to the action of the two-body Green's function at the energy ${E_\Lambda=-\frac{\hbar^2 \Lambda^2}{2\mu}}$ on the ket ${|\delta_\epsilon\rangle}$:
\begin{equation}
\langle {\mathbf k}| \phi_\epsilon^\Lambda \rangle = - \frac{2\mu}{\hbar^2} \times \frac{\chi_\epsilon(k)}{k^2 + \Lambda^2 } .
\end{equation}
Equation~\eqref{eq:r_espilon_subtractive} illustrates the fact that the $\Lambda$ potential is a regularization techniques based on a subtractive scheme.

Using the operator ${r^\Lambda_\epsilon}$, the Bethe Peierls contact condition in Eq.~\eqref{eq:contact_D} can be written without referring to any representation as:
\begin{equation}
\lim_{\epsilon \to 0} r^\Lambda_\epsilon \left[\langle \delta_\epsilon | \psi_\epsilon \rangle \right] = \frac{\Omega_D \hbar^2}{2\mu} \frac{A_\psi}{T_D(E_\Lambda)} .
\label{eq:contact_gene}
\end{equation}
Inserting the closure relation in the left hand side of Eq.~\eqref{eq:contact_gene}, the contact condition can be expressed in the momentum representation as:
\begin{equation}
\lim_{\epsilon \to 0}  \int \frac{d^D{k}}{(2\pi)^D} \chi_\epsilon(k) \left[
\langle {\mathbf k} | \Psi_\epsilon \rangle + \frac{\Omega_D A_{\psi} \chi_\epsilon(k)}{ k^2+\Lambda^2} \right]
=  \frac{-A_{\psi}}{f_D(i\Lambda)} 
\label{eq:contact_k}
\end{equation}
where ${\Lambda>0}$, ${f_D(i\Lambda)=-2 \mu T_D(E_\Lambda)/(\Omega_D \hbar^2)}$ is the scattering amplitude in the Bethe Peierls model evaluated at negative energy and 
\begin{equation}
A_\Psi= \frac{-1}{\Omega_D} \lim_{k\to \infty} \left( k^2 \lim_{\epsilon\to 0} \langle {\mathbf k} | \Psi_\epsilon \rangle\right) .
\end{equation}
For example, in the ${\Lambda-\epsilon}$ potential model a scattering state of incoming momentum ${\mathbf k_0}$ reads:
\begin{equation}
\langle {\mathbf k} | \psi_\epsilon \rangle = (2\pi)^3 \delta(\mathbf k-\mathbf k_0) + \frac{\Omega_D A_\psi \chi_\epsilon(k)}{k_0^2 + i 0^+ - k^2} 
\label{eq:fk_A_Psi}
\end{equation}
where ${A_\psi=-f_D(k_0)}$.

\subsection{Link with the renormalization of the coupling constant}

In what follows, Eq.~\eqref{eq:contact_k} is derived by using the model potential:
\begin{equation}
V(\mathbf r)= v_Q  g_Q(\mathbf r) ,
\end{equation}
where ${v_Q}$ is the bare coupling constant and ${g_Q(\mathbf r)}$ is a family of functions converging toward a $\delta$-distribution in the limit of arbitrarily high values of the ultraviolet momentum cutoff $Q$ ${(Q \to \infty)}$. For a two-body scattering state with an incoming plane wave of  momentum ${|\mathbf k_0\rangle}$, the transition matrix of the model is
\begin{equation}
\langle \mathbf k |T(E+i0^{+}) | \mathbf k_0 \rangle = \langle \mathbf k | V |\Psi_{\mathbf k_0}\rangle .
\end{equation}
The Fourier transform of the function $g_Q(\mathbf r)$ is denoted by
\begin{equation}
\tilde{g}_Q(\mathbf k) =  \int d^3r \exp(-i\mathbf k \cdot \mathbf r ) g_Q(\mathbf r)
\end{equation}
so that the transition matrix reads:
\begin{equation}
\langle \mathbf k |T(E+i0^{+}) | \mathbf k_0 \rangle = v_Q \int \frac{d^3k'}{(2\pi)^3} \tilde{g}_Q(\mathbf k-\mathbf k')  \langle \mathbf k' | \Psi_{{\bf k_0}}\rangle . 
\label{eq:Schrobis}
\end{equation}
A simple choice for the function ${\tilde{g}_Q}$ is:
\begin{equation}
 \tilde{g}_Q =\left\{
\begin{array}{ll}
1 & \mbox{if} \ k<Q\\
0 & \mbox{otherwise}
\end{array}
\right. .
\end{equation}
In the limit of small momenta as compared to the ultraviolet momentum cutoff $Q$ (${|\mathbf k |\ll Q}$ and  ${|\mathbf k_0|\ll Q}$) or equivalently in the limit of large ultraviolet momentum cutoff, the following identity holds:
\begin{equation}
{\int \frac{d^3k'}{(2\pi)^3} \tilde{g}_Q(\mathbf k-\mathbf k')  \langle \mathbf k' | \Psi_{\mathbf k_0} \rangle \operatornamewithlimits{=}_{Q\to \infty} 
\int_{k'<Q} \frac{d^3k'}{(2\pi)^3}  \langle \mathbf k' | \Psi_{\mathbf k_0} \rangle} ,
\end{equation}
which makes possible to directly identify the scattering amplitude of the model at low collisional energy in Eq.~\eqref{eq:Schrobis}:
\begin{multline}
f_Q(\mathbf k_0) = - \frac{\mu}{2\pi \hbar^2}\langle \mathbf k_0 |T(E+i0^{+}) | \mathbf k_0 \rangle \\
\operatornamewithlimits{=}_{Q \to \infty}  -\frac{\mu v_Q}{2\pi \hbar^2} \int_{k'<Q} \frac{d^3k'}{(2\pi)^3}  \langle \mathbf k' | \Psi_{\mathbf k_0} \rangle .
\label{eq:tmatrix-vQ}
\end{multline}
Comparing with Eq.~\eqref{eq:fk_A_Psi} one obtains the relation:  
\begin{equation}
A_{\Psi}  = \lim_{Q \to \infty} \frac{\mu v_Q}{2\pi\hbar^2} \int_{k'<Q} \frac{d^3k'}{(2\pi)^3}  \langle \mathbf k' | \Psi_{\mathbf k_0} \rangle .
\label{eq:magic}
\end{equation}
Using the fact that 
\begin{equation}
\langle {\mathbf k} | \Psi_{\mathbf k_0} \rangle = (2\pi)^3 \delta(\mathbf k-\mathbf k_0) + \frac{\langle {\mathbf k} |T(E+i0^{+}) | \mathbf k_0 \rangle}{E + i 0^+ -\frac{\hbar^2 k^2}{2\mu}} ,
\label{eq:def_T}
\end{equation}
one obtains a closed equation for ${f_Q(\mathbf k_0)}$:
\begin{multline}
-\frac{2\pi \hbar^2}{\mu v_Q} f_Q(\mathbf k_0) 
\operatornamewithlimits{=}_{Q\to \infty} 1 - 4\pi f_Q(\mathbf k_0)  \\ \times \int_{k'<Q} \frac{d^3k'}{(2\pi)^3} \frac{1}{k_0^2 +i0^+ -k^2} .
\end{multline}
The integral in the right hand side of this last equation is performed easily by using the standard analytical continuation ${k_0=iq}$ (where ${q>0}$) and finally:
\begin{equation}
f_Q(\mathbf k_0) \operatornamewithlimits{=}_{Q\to \infty}  - \frac{1}{\frac{2\pi\hbar^2}{\mu v_Q}+\frac{2Q}{\pi} + ik_0} .
\end{equation}
The value of the bare coupling constant ${v_Q}$ is chosen such that the scattering amplitude ${f_Q(\mathbf k_0)}$ coincides with the Bethe Peierls scattering amplitude ${f_3(k_0)}$ in the limit of an arbitrarily large momentum cutoff $Q$, thus yielding:
\begin{equation}
 \frac{1}{a_3} \operatornamewithlimits{=}_{Q\to \infty} \frac{2\pi\hbar^2}{\mu v_Q}+\frac{2Q}{\pi} .
\label{eq:ren}
\end{equation}
Using the fact that ${|a_3|\ll Q}$, the bare coupling constant can thus be expressed as:
\begin{equation}
 v_Q \operatornamewithlimits{=}_{Q\to \infty} - \frac{\pi^2 \hbar^2}{\mu Q} - \frac{\pi^3 \hbar^2}{2\mu a_3 Q^2} +\dots
\end{equation}
From Eqs.~\eqref{eq:ren} and \eqref{eq:magic} one obtains:
\begin{equation}
\lim_{Q\to \infty} \int_{k<Q} \frac{d^3k}{(2\pi)^3}  \left[ \langle \mathbf k | \Psi_{\mathbf k_0} \rangle +\frac{4\pi  A_{\Psi}}{k^2}\right] = \frac{A_{\Psi}}{a_3} .
\label{eq:contact_Q}
\end{equation}
Finally, the identity
\begin{equation}
\lim_{Q\to \infty} \int_{k<Q} \frac{d^3k}{(2\pi)^3} \left( \frac{4\pi}{k^2}-\frac{4\pi}{k^2+\Lambda^2} \right) = \Lambda  \quad (\Lambda \ge 0)
\end{equation}
permits to rewrite Eq.~\eqref{eq:contact_Q} in the same form as the contact condition of Eq.~\eqref{eq:contact_k}.


\begin{thebibliography}{0}
\expandafter\ifx\csname natexlab\endcsname\relax\def\natexlab#1{#1}\fi
\expandafter\ifx\csname bibnamefont\endcsname\relax
  \def\bibnamefont#1{#1}\fi
\expandafter\ifx\csname bibfnamefont\endcsname\relax
  \def\bibfnamefont#1{#1}\fi
\expandafter\ifx\csname citenamefont\endcsname\relax
  \def\citenamefont#1{#1}\fi
\expandafter\ifx\csname url\endcsname\relax
  \def\url#1{\texttt{#1}}\fi
\expandafter\ifx\csname urlprefix\endcsname\relax\def\urlprefix{URL }\fi
\providecommand{\bibinfo}[2]{#2}
\providecommand{\eprint}[2][]{\url{#2}}

\end{thebibliography}


\begin{thebibliography}{99}

\bibitem{Ols98} M. Olshanii, 
{Phys. Rev. Lett. {\bf 81}, 938 (1998)}.

\bibitem{Pet00} D.S. Petrov, M. Holzmann, and G.V. Shlyapnikov, 
{Phys. Rev. Lett. {\bf 84}, 2551 (2000)}.

\bibitem{Pet01} D.S. Petrov and G.V. Shlyapnikov, 
{Phys. Rev. A {\bf 64}, 012706 (2001)}.

\bibitem{Pet04a} D.S. Petrov, C. Salomon, and G.V. Shlyapnikov, 
{Phys. Rev. Lett. {\bf 93}, 090404 (2004)}.

\bibitem{Abr89} A.A. Abrikosov, L.P. Gorkov, and I.E. Dzyaloshinskii, {\sl 'Methods of Quantum Field Theory in Statistical Physics'}, Dover, New York (1975).

\bibitem{Bla86} J.P. Blaizot and G. Ripka, {\sl 'Quantum Theory of Finite Systems'}, The MIT Press, Cambridge Mass. (1986).

\bibitem{Fet03} A.L. Fetter and J.D. Walecka, {\sl 'Quantum Theory of Many-Particle Systems'}, Dover, New York  (2003).

\bibitem{Hua57} K. Huang and C. N. Yang, 
{Phys. Rev. {\bf 105}, 767 (1957)}.

\bibitem{Pri00} L. Pricoupenko, 
{\sl arXiv:cond-mat/0006263}.

\bibitem{Ols01} M. Olshanii and L. Pricoupenko, Phys. Rev. Lett. {\bf 88}, 010402 (2001).

\bibitem{Hug59} N. Hugenholtz and D. Pines, 
{Phys. Rev. {\bf 116}, 489 (1959)}.

\bibitem{Pri04b} L. Pricoupenko, Phys. Rev. A {\bf 70}, 013601 (2004).


\bibitem{Pop72} V.N. Popov, Theor. Math. Phys. {\bf 11}, 565 (1972).

\bibitem{Mor03} C. Mora and Y. Castin, 
{Phys. Rev. A {\bf 67}, 053615 (2003)}.

\bibitem{Pil05} S. Pilati, J. Boronat, J. Casulleras, and S. Giorgini, 
{Phys. Rev. A {\bf 71}, 023605 (2005)}.

\bibitem{Ast09} G.E. Astrakharchik, J. Boronat, J. Casulleras, I.L. Kurbakov, Yu.E. Lozovik, 
{Phys. Rev. A {\bf 79}, 051602(R) (2009)}.

\bibitem{Mor09} C. Mora and Y. Castin, 
{Phys. Rev. Lett. {\bf 102}, 180404 (2009)}.

\bibitem{Gir59} M. Girardeau, and R. Arnowitt, 
{Phys. Rev. {\bf 113}, 755 (1959)}.

\bibitem{Gri96} A. Griffin, 
{Phys. Rev. B {\bf 53}, 9341 (1996)}.

\bibitem{Iso99} T. Isoshima and K. Machida, 
{Phys. Rev. A {\bf 59}, 2203 (1999)} 

\bibitem{Pri11} L. Pricoupenko, Phys. Rev. A {\bf 83}, 062711 (2011).

\bibitem{Che01} A.Yu. Cherny and A. A. Shanenko, 
{Phys. Rev. E {\bf 64}, 027105 (2001)}.

\bibitem{Pri07b} L. Pricoupenko and Y. Castin, J. Phys. A: Math. Theor. {\bf 40}, 12863 (2007).


\end{thebibliography}
\end{document}